\documentclass[12pt]{article}
\usepackage{axodraw}

\catcode`@=12
\begin{document}
\thispagestyle{empty}
\begin{flushright} 
UCRHEP-T353\\ 
March 2003\
\end{flushright}
\vspace{0.5in}
\begin{center}
{\LARGE	\bf Form Invariance of the Neutrino Mass Matrix\\}
\vspace{1.5in}
{\bf Ernest Ma\\}
\vspace{0.2in}
{\sl Physics Department, University of California, Riverside, 
California 92521\\}
\vspace{1.5in}
\end{center}
\begin{abstract}\
Consider the most general $3 \times 3$ Majorana neutrino mass matrix $\cal M$. 
Motivated by present neutrino-oscillation data, much theoretical effort is 
directed at reducing it to a specific texture in terms of a small number of 
parameters.  This procedure is often {\it ad hoc}.  I propose instead that 
for any $\cal M$ one may choose, it should satisfy the condition $U {\cal M} 
U^T = {\cal M}$, where $U \neq 1$ is a specific unitary matrix such that 
$U^N$ represents a well-defined discrete symmetry in the $\nu_{e,\mu,\tau}$ 
basis, $N$ being a particular integer not necessarily equal to one. I 
illustrate this idea with a number of examples, including the 
realistic case of an inverted hierarchy of neutrino masses.
\end{abstract}
\newpage
\baselineskip 24pt

Atmospheric neutrino oscillations have been firmly established \cite{atm} 
now for more than 2 years.  Solar neutrino oscillations have also recently 
been confirmed \cite{sol}.  The atmospheric mixing angle is maximal or 
nearly so with $\Delta m^2 \sim 2.5 \times 10^{-3}$ eV$^2$, whereas the 
solar mixing angle is not maximal but large $(\tan^2 \theta \sim 0.45)$ 
with 2 solutions for $\Delta m^2$, one on either side of $10^{-4}$ eV$^2$.
Together, the neutrino mixing matrix is now determined to a very good first 
approximation by
\begin{equation}
\pmatrix {\nu_1 \cr \nu_2 \cr \nu_3} = \pmatrix {\cos \theta & \sin \theta/
\sqrt 2 & \sin \theta/\sqrt 2 \cr -\sin \theta & \cos \theta/\sqrt 2 & 
\cos \theta/\sqrt 2 \cr 0 & -1/\sqrt 2 & 1/\sqrt 2} \pmatrix {\nu_e \cr 
\nu_\mu \cr \nu_\tau},
\end{equation}
where $\nu_{1,2,3}$ are neutrino mass eigenstates.  In the above, 
$\sin^2 2 \theta_{atm} = 1$ is already assumed and $\theta$ is the solar 
mixing angle.  The $U_{e3}$ entry has been assumed zero but it is only 
required to be small \cite{react}, i.e. $|U_{e3}| < 0.16$. 

It is the aim of much theoretical effort in the past several years 
\cite{review} to find the correct neutrino mass matrix which will fit 
all the data.  The starting point is usually the assumption that there 
are only 3 neutrinos and that they are Majorana fermions.  The most 
general neutrino mass matrix in the basis $\nu_{e,\mu,\tau}$ (where the 
charged-lepton mass matrix is diagonal) is then of the form
\begin{equation}
{\cal M} = \pmatrix {A & D & E \cr D & B & F \cr E & F & C},
\end{equation}
where $A,B,C$ may be chosen real by redefining the phases of 
$\nu_{e,\mu,\tau}$, but then $D,E,F$ remain complex in general.  Any model of 
neutrino mass (of which there are very many in the literature) always ends up 
with a simplification of $\cal M$, thereby reducing the number of independent 
parameters.  The resulting form of $\cal M$ is of course always chosen to be 
consistent with experimental data, so that the model may be declared a 
success.  This procedure is sometimes rather {\it ad hoc} and rife with 
arbitrary assumptions.  Instead, I propose below a \underline {novel} 
approach based on symmetry arguments.

Consider a \underline {specific} unitary transformation $U$.  Let $\nu'_i = 
U_{ij} \nu_j$, then $\cal M$ becomes $U {\cal M} U^T$ in the $\nu'$ basis. 
I propose that
\begin{equation}
U {\cal M} U^T = {\cal M}
\end{equation}
be required as a condition on $\cal M$.  If $U$ represents a well-defined 
discrete symmetry, then this is nothing new.  However, Eq.~(3) also implies 
that
\begin{equation}
U^n {\cal M} (U^T)^n = {\cal M},
\end{equation}
where $n$ = 1,2,3, etc.  This sequence should terminate at $n = \bar n$ with 
$U^{\bar n} = 1$.  Otherwise, the only possible solution for $\cal M$ would 
be a multiple of the identity matrix (in the case that $U$ is also real).  
My proposal is that for a particular value $N < \bar n$, $U^N$ should 
represent a well-defined discrete symmetry in the $\nu_{e,\mu,\tau}$ basis.  
Again if $N = 1$, there is nothing new.  However, if $N \neq 1$, say 2, then 
the unitary matrix $U$ in Eq.~(3) represents rather the ``square root'' of 
the discrete symmetry $U^2$.  This is a new idea, with very interesting 
consequences as shown below.

Consider first the simple discrete symmetry
\begin{equation}
\nu_e \to \nu_e, ~~~ \nu_{\mu,\tau} \to - \nu_{\mu,\tau},
\end{equation}
i.e.
\begin{equation}
U = \pmatrix {1 & 0 & 0 \cr 0 & -1 & 0 \cr 0 & 0 & -1}.
\end{equation}
The requirement of Eq.~(3) fixes $D = E = 0$, thus
\begin{equation}
{\cal M} = \pmatrix {A & 0 & 0 \cr 0 & B & F \cr 0 & F & C}.
\end{equation} 
Now suppose instead that
\begin{equation}
U^2 = \pmatrix {1 & 0 & 0 \cr 0 & -1 & 0 \cr 0 & 0 & -1}.
\end{equation}
Then there are two obvious solutions for $U$, i.e.
\begin{equation}
U_1 = \pmatrix {1 & 0 & 0 \cr 0 & i & 0 \cr 0 & 0 & i}, ~~~ 
U_2 = \pmatrix {1 & 0 & 0 \cr 0 & i & 0 \cr 0 & 0 & -i},
\end{equation}
resulting in
\begin{equation}
{\cal M}_1 = \pmatrix {A & 0 & 0 \cr 0 & 0 & 0 \cr 0 & 0 & 0}, ~~~ 
{\cal M}_2 = \pmatrix {A & 0 & 0 \cr 0 & 0 & F \cr 0 & F & 0}.
\end{equation}
Note that ${\cal M}_1$ and ${\cal M}_2$ are both special cases of the 
$\cal M$ of Eq.~(7).  Note also that ${\cal M}_2$ may be obtained in general 
with $U_2$ of the form
\begin{equation}
U_2 = \pmatrix {1 & 0 & 0 \cr 0 & e^{2 \pi i/n} & 0 \cr 0 & 0 & 
e^{-2 \pi i/n}},
\end{equation}
where $n \geq 3$.  This means that $U_2$ itself already represents a 
well-defined discrete symmetry in the $\nu_{e,\mu,\tau}$ basis, and there is 
nothing new about this application.  On the other hand, neither ${\cal M}_1$ 
nor ${\cal M}_2$ are realistic candidates for the neutrino mass matrix.

Consider next the simple interchange discrete symmetry
\begin{equation}
\nu_e \to \nu_e, ~~~ \nu_\mu \leftrightarrow \nu_\tau,
\end{equation}
i.e.
\begin{equation}
U = \pmatrix {1 & 0 & 0 \cr 0 & 0 & 1 \cr 0 & 1 & 0}.
\end{equation}
The requirement of Eq.~(3) fixes $D = E$ and $B = C$, thus
\begin{equation}
{\cal M} = \pmatrix {A & D & D \cr D & B & F \cr D & F & B}.
\end{equation}
This is now a very good candidate for a realistic neutrino mass matrix. 
In fact, if the four parameters $A,B,D,F$ are chosen real, then this 
$\cal M$ is exactly diagonalized with Eq.~(1).  It is also the form 
advocated recently \cite{allp} as an all-purpose neutrino mass, where 
it is written as
\begin{equation}
{\cal M} = \pmatrix {a+2b+2c & d & d \cr d & b & a+b \cr d & a+b & b}.
\end{equation}
Depending on the actual values of $a,b,c,d$, this $\cal M$ was shown to have 
7 different solutions, 3 corresponding to the normal hierarchy, 2 to an 
inverted hierarchy, and 2 to three nearly degenerate neutrino masses.  
However, the symmetry of Eq.~(12) cannot choose among these 7 solutions.

Specific examples of Eq.~(14) which have appeared in the literature include
the cases $A=B+F$ \cite{monu}, $A+D=B+F$ \cite{hps}, and $A+B+F=0$ 
\cite{josh}. It should also be pointed out that a \underline {complete} 
\underline {theory} exists for 3 nearly degenerate neutrino masses where 
the observed ${\cal M}_\nu$ is derived from a radiatively corrected \cite{bmv} 
neutrino mass matrix based on the discrete symmetry $A_4$ \cite{a4}.  In this 
model, the parameters $b,c,d$ of Eq.~(15) are generated in one-loop order by 
new physics at the TeV scale.  This implies that the 
effective mass $m_0$ observed in neutrinoless double beta decay 
\cite{klapdor} should not be too small, or else the interpretation of 
$\Delta m^2 \sim 2.5 \times 10^{-3}$ eV$^2$ for atmospheric neutrino 
oscillations as a radiative correction becomes rather unnatural.  With 
the recent data from the Wilkinson Microwave Anisotropy Probe (WMAP), this 
mass also gets an upper bound \cite{wmap} of 0.23 eV.  The radiative $A_4$ 
model would require $m_0$ to be observable with some experimental 
improvement in either neutrinoless double beta decay or WMAP.

Now suppose instead that
\begin{equation}
U^2 = \pmatrix {1 & 0 & 0 \cr 0 & 0 & 1 \cr 0 & 1 & 0}.
\end{equation}
Then one obvious solution of its square root is
\begin{equation}
U_1 = \pmatrix {1 & 0 & 0 \cr 0 & (1-i)/2 & (1+i)/2 \cr 0 & (1+i)/2 & 
(1-i)/2}, 
\end{equation}
resulting in
\begin{equation}
{\cal M}_1 = \pmatrix {A & D & D \cr D & B & B \cr D & B & B} 
= \pmatrix {2b+2c & d & d \cr d & b & b \cr d & b & b}, 
\end{equation}
in the notations of Eqs.~(14) and (15), where $F=B$ and $a=0$ have now been 
fixed respectively. The mass eigenvalues are then
\begin{equation}
m_{1,2} = 2b+c \mp \sqrt {c^2 + 2d^2}, ~~~ m_3 = 0.
\end{equation}
Since $m_3$ corresponds to the mass eigenstate $\nu_3 = (\nu_\mu - \nu_\tau)/
\sqrt 2$, this solution is an inverted hierarchy with
\begin{eqnarray}
(\Delta m^2)_{atm} &\simeq& (2b+c)^2 \simeq 4b^2, \\ 
(\Delta m^2)_{sol} &\simeq& 4(2b+c) \sqrt {c^2+2d^2} \simeq 8b 
\sqrt {c^2+2d^2}.
\end{eqnarray}
Another solution is not so obvious, namely
\begin{equation}
U_2 = {1 \over \sqrt 3} \pmatrix {1 & 1 & 1 \cr 1 & \omega & \omega^2 \cr 
1 & \omega^2 & \omega}, 
\end{equation}
with $\omega = e^{2 \pi i/3}$, resulting in
\begin{equation}
{\cal M}_2 = \pmatrix {2b+2d & d & d \cr d & b & b \cr d & b & b},
\end{equation}
which is a reduction of ${\cal M}_1$ of Eq.~(18) by the condition $c=d$, thus 
predicting
\begin{equation}
\tan^2 \theta = 2 - \sqrt 3 = 0.27,
\end{equation}
which is marginally allowed by present experimental data at the low end of 
an acceptable range of values.

Whereas $U^2$ of Eq.~(16) is the realization of the simple interchange 
discrete symmetry of Eq.~(12), both $U_1$ of Eq.~(17) and $U_2$ of Eq.~(22) 
are not.  Note however that the eigenvalues of $U^2$ 
are (1,1,--1), whereas those of $U_1$ and $U_2$ are $(1,1,-i)$ and $(1,-1,+i)$ 
respectively.

Another possible choice of a simple discrete symmetry is
\begin{equation}
U = \pmatrix {0 & 1 & 0 \cr 0 & 0 & 1 \cr 1 & 0 & 0},
\end{equation}
which results in
\begin{equation}
{\cal M} = \pmatrix {A & D & D \cr D & A & D \cr D & D & A}.
\end{equation}
Now
\begin{equation}
U^2 = \pmatrix {0 & 0 & 1 \cr 1 & 0 & 0 \cr 0 & 1 & 0}
\end{equation}
also results in the same $\cal M$ and $U^3 = 1$ with the eigenvalues of both 
$U$ and $U^2$ being $(1, \omega, \omega^2)$.  However this $\cal M$ is not 
a realistic candidate for the neutrino mass matrix.

Going back to Eq.~(23), we see that $d$ has to be much smaller than $b$ to 
explain $(\Delta m^2)_{sol} << (\Delta m^2)_{atm}$.  Suppose we now set $d=0$, 
then ${\cal M}_2$ has a two-fold degeneracy, i.e. $m_{1,2} = 2b$, $m_3 = 0$, 
with maximal $\nu_\mu - \nu_\tau$ mixing.  This is then a very good starting 
point also for the understanding of solar neutrino oscillations in terms of 
an inverted hierarchy where $(\Delta m^2)_{sol}$ and the solar mixing angle 
$\theta$ are radiative effects, in analogy to that of Ref.~[9].

Consider thus the most general radiative corrections to ${\cal M}_\nu$, i.e.
\begin{equation}
R = \pmatrix {r_{ee} & r_{e\mu} & r_{e\tau} \cr r^*_{e\mu} & r_{\mu\mu} & 
r_{\mu\tau} \cr r^*_{e\tau} & r^*_{\mu\tau} & r_{\tau\tau}},
\end{equation}
then
\begin{equation}
{\cal M}_\nu \to (1 + R) {\cal M}_\nu (1 + R^T),
\end{equation}
and becomes
\begin{eqnarray}
m_0 \pmatrix {1+2r_{ee} & r^*_{e\mu} + (r_{e\mu}+r_{e\tau})/2 & 
r^*_{e\tau} + (r_{e\mu}+r_{e\tau})/2 \cr r^*_{e\mu} + (r_{e\mu}+r_{e\tau})/2 
& (1+2r_{\mu\mu}+2r_{\mu\tau})/2 & (1+r_{\mu\mu}+r_{\tau\tau}+r_{\mu\tau}+
r^*_{\mu\tau})/2 \cr r^*_{e\tau} + (r_{e\mu}+r_{e\tau})/2 & (1+r_{\mu\mu}+
r_{\tau\tau}+r_{\mu\tau}+r^*_{\mu\tau})/2 & (1+2r_{\tau\tau}+
2r^*_{\mu\tau})/2}
\end{eqnarray}

Let
\begin{eqnarray}
c &\equiv& r_{\mu\mu}+r_{\tau\tau}+2Re(r_{\mu\tau}) - 2r_{ee} > 0, \\ 
d &\equiv& \sqrt 2 Re(r_{e\mu}+r_{e\tau}),
\end{eqnarray}
then the mass eigenvalues of the radiatively corrected ${\cal M}_\nu$ are
\begin{eqnarray}
m_{1,2} = \left[ 1 + 2r_{ee} + {c \over 2} \mp {1 \over 2} \sqrt {c^2+4d^2} 
\right] m_0, ~~~  m_3 = O(r^2) ~m_0,
\end{eqnarray}
with the solar mixing angle $\theta$ given by
\begin{equation}
\tan^2 \theta = 1 - {2c \over \sqrt {c^2+4d^2} + c},
\end{equation}
and
\begin{equation}
U_{e3} \simeq {1 \over \sqrt 2} (r_{e\mu} - r_{e\tau}).
\end{equation}

In the Standard Model, $r_{ij}=0$ for $i \neq j$, i.e. $R$ is diagonal, hence 
$d=0$ and even though $m_1$ and $m_2$ are split because $c \neq 0$, there is 
no mixing so $\nu_e$ does not oscillate at all.  To obtain solar neutrino 
oscillations, we need flavor-changing interactions.  As a simple example, 
consider the addition of 3 charged scalar singlets $\chi^+_{1,2,3}$ with 
the following interactions:
\begin{eqnarray}
{\cal L}_{int} &=& f [ (\nu_1 l_3 - l_1 \nu_3) \chi^+_1 + (\nu_2 l_3 - l_2 
\nu_3) \chi^+_2 ] \nonumber \\ &+& h (\nu_1 l_2 - l_1 \nu_2) 
\chi^+_3 + H.c. + m^2_{ij} \chi^+_i \chi^-_j,
\end{eqnarray}
where $l_1 = e$, $l_{2,3} = (\mu \pm \tau)/\sqrt 2$, and correspondingly for 
the neutrinos.  This Lagrangian is invariant under the discrete symmetry
\begin{eqnarray}
(\nu_1,l_1) \leftrightarrow (\nu_2,l_2), &~& (\nu_3,l_3) \to (\nu_3,l_3), 
\nonumber \\ \chi^+_1 \leftrightarrow \chi^+_2, &~& \chi^+_3 \to -\chi^+_3,
\end{eqnarray}
which is broken softly by $m^2_{ij}$.

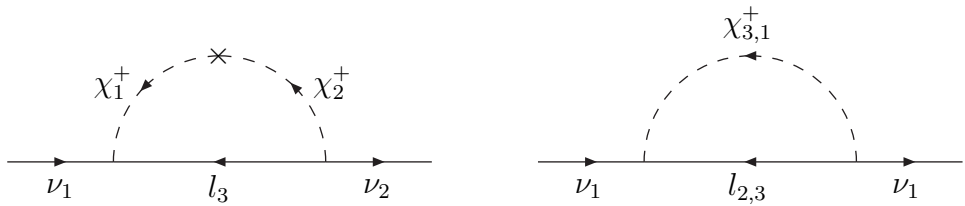
\begin{figure}[htb]
\begin{center}
\begin{picture}(360,80)(0,0)
\ArrowLine(0,10)(40,10)
\ArrowLine(120,10)(40,10)
\ArrowLine(120,10)(160,10)
\ArrowLine(200,10)(240,10)
\ArrowLine(320,10)(240,10)
\ArrowLine(320,10)(360,10)
\DashArrowArc(80,10)(40,90,180){4}
\DashArrowArc(80,10)(40,0,90){4}
\DashArrowArc(280,10)(40,0,180){4}

\Text(20,0)[]{$\nu_1$}
\Text(80,0)[]{$l_3$}
\Text(140,0)[]{$\nu_2$}
\Text(220,0)[]{$\nu_1$}
\Text(280,0)[]{$l_{2,3}$}
\Text(340,0)[]{$\nu_1$}
\Text(80,50)[]{$\times$}
\Text(40,40)[]{$\chi_1^+$}
\Text(123,40)[]{$\chi_2^+$}
\Text(280,62)[]{$\chi^+_{3,1}$}
\end{picture}
\end{center}
\caption{Neutrino wave-function renormalizations.}
\end{figure}

The radiative corrections $r_{ij}$ are easily calculated in one loop as 
shown in Fig.~1.  Note that $c$ and $d$ of Eqs.~(31) and (32) are 
\underline {finite} and derivable from the parameters of Eq.~(36).
In the inverted hierarchy of neutrino masses, 
$m_0 = \sqrt {(\Delta m^2)_{atm}}$, hence
\begin{equation}
2 \sqrt {c^2+4d^2} = {(\Delta m^2)_{sol} \over (\Delta m^2)_{atm}} 
\simeq 0.04,
\end{equation}
and for $\tan^2 \theta = 0.45$, $d/c = 1.22$.  In this model, let
$\chi_i = V_{ij} \chi'_j$, where $\chi'_j$ are mass eigenstates with 
masses $m_j$, then
\begin{eqnarray}
c &=& {f^2 \over 16 \pi^2} \sum_{j=1}^3 (|V_{2j}|^2 - |V_{1j}|^2) 
\ln m_j^2, \\ d &=& {f^2 \over 16 \pi^2} Re \sum_{j=1}^3 V^*_{2j} V_{1j} 
\ln m_j^2, \\ U_{e3} &=& {-fh \over 32 \pi^2} \sum_{j=1}^3 V^*_{2j} V_{3j} 
\ln m_j^2.
\end{eqnarray}
Realistic values for $c$ and $d$ as well as a nonnegligible complex $U_{e3}$ 
are then possible if $f$ and $h$ are of order unity, and the $m_j$'s are 
sufficiently different from one another.

Flavor-changing leptonic decays are predicted.  For example, the amplitude 
for $\mu \to e \gamma$ is given by
\begin{equation}
{\cal A} = {e f m_\mu \over 768 \pi^2} \sum_{j=1}^3 (f V_{1j}^* - h V_{3j}^*) 
{V_{2j} \over m_j^2} \epsilon^\lambda q^\nu \bar e \sigma_{\lambda\nu} 
(1+\gamma_5) \mu,
\end{equation}
whereas that of $\tau \to e \gamma$ is obtained by replacing $m_\mu$ by 
$m_\tau$ and $h$ by $-h$.  This means that one or the other of these 
decays may be suppressed but not both.  Masses for $\chi'_j$ of order 1 TeV 
are consistent with the present experimental upper bounds on the 
corresponding branching fractions.

In conclusion, a form invariance of the neutrino mass matrix has been 
proposed, i.e. $U {\cal M}_\nu U^T = {\cal M}_\nu$, where $U$ is a specific 
unitary matrix and $U^N$ (with $N$ not necessarily equal to one) represents 
a well-defined discrete symmetry in the $\nu_{e,\mu,\tau}$ basis.  Using 
Eq.~(12) as the definition of $U^2$, Eqs.~(18) and (23) have been derived, 
allowing for an inverted hierarchy of neutrino masses, suitable for 
explaining the present data on solar and atmospheric neutrino oscillations. 
The possible radiative origin of $(\Delta m^2)_{sol}$, $\tan^2 \theta_{sol}$, 
as well as $U_{e3}$ has also been shown in a simple specific model with new 
flavor-changing interactions.\\

This work was supported in part by the U.~S.~Department of Energy
under Grant No.~DE-FG03-94ER40837.\\

\bibliographystyle{unsrt}

\end{document}